\documentclass[twocolumn,runningheads]{svjour2}
\journalname{Astrophysics and Space Science}
\smartqed  % flush right qed marks, e.g. at end of proof
\usepackage{graphicx}
\usepackage{mathptmx}      % use Times fonts if available on your TeX system
\usepackage{natbib}
\usepackage {ifthen}
\def\gray {$\gamma$-ray\ }
\def\grays{$\gamma$-rays\ }

\def\deg{^\circ}

\def\regionA{$330\deg<l<30\deg,\ |b|<5\deg$}
\def\regionH{$300\deg<l<60\deg,\   |b|<10\deg$  }

\def\Lg{L_\gamma}
\def\Lgunits{ph s$^{-1}$\ }
\def\Lgergunits{erg\ s$^{-1}$\ }
\def\fluxunits{cm$^{-2}$  s$^{-1}$\ }
\def\Edot{\dot{E}}

\def\apjl{ApJL}    %AWS
\def\aap {A\&A }  %AWS    
\begin{document}
\title{Source population synthesis and the Galactic diffuse gamma-ray emission }

   \subtitle{ }

   \author{A. W. Strong  
          }

%\offprints{A. W. Strong, aws@mpe.mpg.de}

\institute{ Max-Planck-Institut f\"ur extraterrestrische Physik,
Postfach 1312, D-85741 Garching, Germany           
              \email{aws@mpe.mpg.de}
             }

%\date{  DRAFT 0.001  Received   / Accepted  }

\titlerunning{Source population synthesis }
\authorrunning{Strong A.W. }
\maketitle

\begin{abstract}

Population synthesis is used to study the contribution from unresolved sources
to the Galactic ridge emission measured by EGRET.
%Both high and low luminosity populations are considered, and a range of luminosity functions
%and spatial distributions are considered.
Synthesized source counts are compared with the 3rd EGRET catalogue at low and high latitudes.
For pulsar-like populations, 5-10\% of the emission $>100$ MeV comes from sources
below the EGRET threshold.
A steeper luminosity function can increase this to  20\%  without violating EGRET source statistics.
 Less luminous populations can produce much higher values
without being detected.
% and we place constraints on their parameters considering limits from
%models of interstellar emission.
Since the unresolved source spectrum is different from the interstellar spectrum,
it could  provide an explanation of the observed MeV and GeV excesses above the predictions, and we give an explicit example
of how this could work.
%Such population studies, already important for EGRET, will be essential for GLAST.
%Fluctuations in the unresolved source emission can provide a test of the models.

\keywords{gamma rays \and diffuse emission \and gamma-ray sources}
%\PACS{First \and Second \and More}
\end{abstract}

%\keywords{Gamma rays: observations -- Galaxy: structure -- ISM: general  -- 
%cosmic rays 
 %  }

%
%________________________________________________________________

\section{Introduction}
The Galactic plane is known to be an intense emitter of X-rays and \grays
from keV to at least 100 GeV. 
At X-ray energies, it has recently been claimed 
\citep{2006A&A...452..169R} %Revnivtsev 
that the 2-10 keV emission can be explained entirely by a population
of weak  sources, mainly CVs.
Above 50 keV, unresolved sources also appear to be required to explain the hard-power law emission observed by INTEGRAL
\citep{2005A&A...444..495S,  %Strong 
      2005ApJ...635.1103B}; %Bouchet
 AXPs and/or pulsars are potential candidates.
 This   raises the question of the situation
for \grays, in particular the range observed by EGRET, 30 MeV - 100 GeV.
\gray telescopes  are relatively  insensitive and reveal only the `tip of 
the iceberg' of the sources, so most of them will  go undetected unless it happens
that only strong sources exist in the Galaxy.
Conventional wisdom is that the source contribution to the unresolved ridge emission is at the few percent 
level,
% and so can be neglected in studies of interstellar emission,
% but this has never been systematically investigated using the
%final  EGRET data.
 The problem of the GeV excess in the diffuse emission
compared to the expected spectrum from interstellar processes
\citep{2004ApJ...613..962S},  %SMR2004
 is however a
hint that the emission may have other  components whose contribution
is also energy-dependent. The failure to explain the 1-30 MeV emission measured by COMPTEL
from interstellar components is another pointer to a source  population.
%More recently, INTEGRAL/SPI results on the ridge emission show a hard 
%power-law component difficult to understand in terms of cosmic-ray interactions
%with interstellar gas and radiation.

%The question arises  whether a population of weak sources can make a significant
%contribution to the ridge  emission.
% Although it is not possible to answer this
%question definitively, quite stringent limits can be set based on available data,
%as we will demonstrate. 
% Note that even detectors with much higher sensitivity
%and resolution (Chandra) are insufficient to resolve the emission
%into the proposed weak sources  in the X-ray range (Revnitsev, also Grindlay etal, Hong et al ApJ Dec20 2005).

This topic has wide ranging implications: for example it has been claimed that the GeV excess 
is a signature of dark matter decay 
\citep{2005A&A...444...51D}%deBoer
\footnote{but see  critique by
\citet{2006JCAP...05..006B}.} %Bergstrom
, but this is only plausible if  all  viable alternatives are
excluded, and source populations provide one important such candidate. 
%%%%%%%%%%%%%%%%%%%%%%%%%%%%%%%%%%%%%%%%%%%%%%%%%%%%%%%%%%%%%%%%%%%%%%%%%%%%%%%%%%%%%%%%%%%%%%%%%%%%%%%%%%%%%%%%%%%%%%%%%

\section{\gray source properties}

The 3rd EGRET Catalogue contains 271 sources, including 66 well-identified extragalactic sources (AGN),
leaving about  200 for the present study. A subset of these may also be extragalactic.
Almost the only definite Galactic identifications are 6 pulsars (+ some other pulsar candidates)  and the Crab nebula,
which have isotropic luminosities $\Lg$($>$100~MeV) 
ranging from   $10^{37}$\Lgunits ( 10$^{33}$ \Lgergunits) (Geminga) to   $10^{39}$\Lgunits (10$^{35}$ \Lgergunits) (Crab pulsar).
%These are summarized in Table 1.
A summary of \gray pulsar observations is given by
 \citet{1999ApJ...516..297T}.
The Crab at 2 kpc is perhaps the most luminous and distant Galactic source detected,
while Geminga at 160 pc the least luminous and nearest. The Vela pulsar is intermediate 
in distance (300 pc) and luminosity $10^{38}$\Lgunits (10$^{34}$ \Lgergunits )  although it is the brightest source on the sky
(8 $10^{-6}$ \fluxunits compared to 2  $10^{-6}$ \fluxunits for the Crab).
Other detected or candidate pulsars also lie in this luminosity range.
Apart from pulsars,
plausible identifications exist e.g. for  the  microquasars LS5039 and LSI +61 303,
% and a source near the Galactic centre,
but these do not significantly help the present investigation.
Most sources with  $\Lg$($>$100 MeV)$<$  10$^{33}$ \Lgergunits  are invisible to us with present instrumentation.
The data are too sparse to construct a luminosity function directly, so we leave this as a  parameterized input.
% The unidentified sources near the plane have a latitude spread of a few degrees, so
%must have comparable distances to the pulsars, with a probable maximum of a few kpc.
%Based on such limited statistics one cannot expect final conclusions,
%but they provide constraints which allow the parameter-space of source populations to be
%significantly limited, and the contributions to the diffuse emission to be estimated.

We approach the problem using population synthesis, comparing the models   both with
source counts and the intensity of diffuse emission
\footnote{We will often refer to unresolved emission generically  as `diffuse' regardless of its true nature}.
Comparison in various sky regions increases the discrimination power of the comparisons.
Thus source counts in high-latitude regions constrain the 
low--luminosity, nearby source population
%(with the uncertainty of the extragalactic population),
while counts in low-latitude regions constrain the high-luminosity sources,
and diffuse emission in low-latitudes constrains both low and high-luminosity populations.

As usual in such studies a major uncertainty is the spatial distribution of sources,
but with plausible assumptions  this can be modelled.
Once the spatial distribution is assumed, the observed (source-produced) gamma-ray sky 
depends only on  the  luminosity function.
% By considering a series of luminosity
%bands, we can constrain the possible spatial density of sources in each band,
%and/or  determine what its contribution to the diffuse emission could be.
 
For the truly diffuse (interstellar)  emission we make use of the {\it galprop} models
\citep{2004ApJ...613..962S}  %SMR2004
.

The present study is designed to be independent of the physical details of the sources,
 making use of, for example, theoretical \gray luminosity functions of pulsars only for guidance.
We do not want to restrict attention to known classes of objects.
The essential input is  only geometry, the inverse square law,
simple luminosity functions,
 the EGRET source catalogue and  EGRET skymap data.

%We distinguish populations which are known to exist since at least one member is observed, 
%and hypothetical populations which can only be indirectly observed via the diffuse emission.

%State 5 questions: 

%1. What do we expect based on the known sources ?

%2. What properties must a population of weak numerous sources to produce a substantial contribution to the diffuse emission

%3. Can Gemingas/Velas with their hard spectra account for the GeV excess ?

%4. What is the situation in COMPTEL and hard X-ray bands ?

%5. What can be expected from GLAST observations.

%We conclude  that a population of Gemingas/Velas could indeed explain the GeV excess, 
%and that much of the diffuse emission could well be from a weak/numerous population which are otherwise invisible.

%%%%%%%%%%%%%%%%%%%%%%%%%%%%%%%%%%%%%%%%%%%%%%%%%%%%%%%%%%%%%%%%%%%%%%%%%%%%%%%%%%%%%%%%%%%%%%%%
\section{Previous studies}
%%%%%%%%%%%%%%%%%%%%%%%%%%%%%%%%%%%%%%%%%%%%%%%%%%%%%%%%%%%%%%%%%%%%%%%%%%%%%%%%%%%%%%%%%%%%%%%%

Most previous population studies have been aimed at deducing the nature of the EGRET
unidentified sources, either generally or in particular as pulsars.
Although the nature of the sources is not our goal here, these studies are still relevant.
A population synthesis study of the 2nd EGRET catalogue 
 \citep{1996A&AS..120C.461K} %Kanbach
 showed that the unidentified sources had  luminosities $>$100 MeV in the range 
$6\ 10^{34} - 3\ 10^{35}  $\Lgergunits, with 700--3400 objects in the Galaxy.
Flux distributions of EGRET unidentified sources were constructed by 
     \citet{2001ICRC....6.2566R} %Reimer
 and \citet{2000Natur.404..363G} %Gehrels
  in attempts to deduce
the properties of unidentified sources, including division by variability and angular distribution.
\citet{2001AIPC..587..663C} %Chen
  made a population synthesis study of the 3EG catalogue, including disk and isotropic source distributions.
%\citet{2004MSAIS...5..191C} %Chen
%made a similar study focussed on the AGILE mission. 

Other studies considered pulsars specifically.
%\citet{2003sath.conf...19C}: Goulds Belt.  %Cheng
%\citet{2003pasb.conf...93C}: pulsar models.%Cheng
%\citet{2004IAUS..218..419L}: Goulds Belt.  %Leung.
%\citet{2001AIPC..587..580G,2002ASPC..271..335H} % Gonthier; Harding
% made  pulsar population studies.
Very detailed physical pulsar population synthesis studies by
 \citet{2002ApJ...565..482G} %Gonthier 2002
 and
 \citet{2004ApJ...604..775G} % Gonthier et al.
 using their polar-cap model were aimed at establishing
the relation between radio and \gray  properties and predicting detections for future missions like GLAST.
\citet{2001ApJ...548L..37H,2003pasb.conf..115G,2004AdSpR..33..571H,2005Ap&SS.297...71G} % Harding&Zhang,Gonthier et al,Harding et al, Gonthier et al. (also Grenier etal ??)
\citet{2004ApJ...608..418C}  %Cheng
 considered the possibility that gamma-ray sources in the
Gould's Belt  are nearby pulsars.
% (although now (Grenier) finds these sources are not valid due to dark gas).

\citet{1998MNRAS.301..841Z} % Zhang and Cheng
 proposed that pulsars can account for the diffuse GeV excess.
\citet{2000A&A...357..957Z} %Zhang Zhang and Cheng 2000
 made a detailed population synthesis based on the outer-gap model, and also studied the properties of 38 low-latitude unidentified sources, proposing an
a correlation with SNR and OB associations.
% can be explained if the sources are pulsars and have a common origin with SN/OB.  
\citet{1997ApJ...476..347Y} %YadigarogluRomani
proposed that the EGRET unidentified sources are compatible with young pulsars,
while in contrast 
\citet{2003A&A...404..163B} % Bhattacharya
claim they are better traced by spiral arms and molecular clouds.

In view of the uncertainty in the nature of the unidentified sources, a flexible approach to modelling is desirable,
as described in this paper.

\section{Population synthesis}
%%%%%%%%%%%%%%%%%%%%%%%%%%%%%%%%%%%%%%%%%%%%%%%%%%%%%%%%%%%%%%%%%

A general-purpose population synthesis code has been written.
Sources are assigned a density $\rho(R,z,L_\gamma)$ and sampled by standard Monte-Carlo techniques.
Oversampling is used to reduce statistical fluctuations.
The density is normalized to $\rho$ at $R = 8.5$ kpc, in units of sources kpc$^{-3}$.
Power-law luminosity functions within given $L_\gamma$ limits can be generated.
The $(R,z)$ source distribution is here based on  pulsars 
\citep{2004IAUS..218..105L} %Lorimer
as representative of \gray sources, but other distributions are also possible and will be
addressed in future work.
The resulting source list is analyzed to generate differential
source counts N(S) and the total emission spectrum both above and below a given detection threshold.
%We compare with the 3EG catalogue, excluding those with AGN identifications.
%A {\it robust} estimate of the flux from below the threshold for a given model is obtained by 
%scaling the observed total flux  from sources above threshold with the
%model flux ratio below/above threshold. {(\it include this ratio and result of this method in the Tables ?).}
%Longitude and latitude profiles can be generated.
The total spectrum is then combined with interstellar emission models from {\it galprop}
\citep{2004ApJ...613..962S}.  %SMR2004.
We use source counts for one energy range ($>100$ MeV) only; it would be preferable to consider
the energy-dependence of N(S) but the spectral information in the available catalogues is limited.
In future (e.g. for GLAST) this will be feasible.
%The former is preferred for the brighter populations, the latter
% for the dimmer  populations.
%We do not depend on a particular source model,
%but pulsars serve as a suitable starting point
%for which the shape of the luminosity function can be estimated  from theory.
A note about beamed sources, in particular pulsars. For the present purpose a populations of randomly-oriented beamed sources
is fully equivalent
 to a population of unbeamed sources with a lower spatial density of sources
 having an isotropic emission.
Therefore we do not explicitly include beaming in our population synthesis.
%(see Fauche-Giguere and Kaspi astroph/0512585 for the same approach).

%Another note concerning source variabilty.
% All \gray sources are variable at some level,
%and especially the hard X-ray sky is highly variable at least at lower energies.
%(see e.g. \citet{2005A&A...444..495S}).%Strong etal
%Many EGRET sources are variable as well.
%However for the present analysis we are interested in the combined emission from whole
%populations of objects, for which only the {\it time-average flux} is relevant.
%We define our luminosity functions to apply to time-averages, and the comparison
%is made with time-averaged catalogues.
% The effect of time-variations will anyway
%preserve the power-law luminosity function shape  (except at the end-points) which we are using.

%The simulated and observed N(S) are presented in logarithmic binned form $\Delta N(log S) / \Delta log S$
%\footnote{only differential N(S) are considered since integral  forms have undesirable statistical properties and obscure the true distribution}.
%For $N(S)\propto S^{-\Gamma}$,$N(log S)\propto S^{-\Gamma+1}$, $dN(log S)/dlog S=-(\Gamma-1)$,
% i.e. the slope in the plots is the differential slope of N(S) minus one
%\footnote{since $N(S)dS=N(log S) d(log S)$, $N(log S)= S N(S)$}.
%In the usual limiting cases, $\Gamma =  2.5$ for uniform 3D geometry, and  2.0 for uniform 2D (disk) geometry. 
%We use  $ \Delta log_{10} S= 0.25$ since this gives a reasonable compromise between resolution and statistics.

%%%%%%%%%%%%%%%%%%%%%%%%%%%%%%%%%%%%%%%%%%%%%%%
\subsection{Pulsar luminosity function}
%%%%%%%%%%%%%%%%%%%%%%%%%%%%%%%%%%%%%%%%%%%%%%%
We use pulsars just as an guide to the   choice of luminosity function.
For this we use the  luminosity as a function of spin-down power $\Edot$: $L_\gamma\propto\Edot^\beta$
 for which a wide spread exists in the literature depending on the model.
The luminosity function can be then estimated as follows:
$N(L_\gamma)={dN\over dL}={dN\over d\Edot}\ {d\Edot\over dL} = {dN\over dt}\ {dt\over d\Edot}\ {d\Edot\over dL}
 \propto  {dt\over d\Edot}\ {d\Edot\over dL}$ for constant birthrate.
Since $\Edot\propto B^2/P^4\propto \dot{P}/P^3$, $d\Edot/dt\propto\Edot/P^2$ and hence
$N(L_\gamma)\propto B^{-1}L_\gamma^{-(1+2\beta)/2\beta}$
According to the polar-cap model of
 \citet{2002ApJ...565..482G} %Gonthier 2002
$\beta\approx {1\over 2}$, so $N(L_\gamma) \propto L_\gamma^{-2}$  
The slot-gap model of 
\citet{2003ApJ...588..430M} %Muslimov and Harding
gives $\beta={1\over 4}$ so $N(L_\gamma) \propto L_\gamma^{-3}$.% their equation 30 and their Fig 3
The outer-gap model of
\citet{2004ApJ...604..317Z} %Zhang
  gives $\beta=0.4 - 1$ so $N(L_\gamma) \propto L_\gamma^{-2.3}- L_\gamma^{-1.5} $.
The  dynamic range of \gray luminosity in these models is  about 1000.

\ifthenelse{0=1}
{
%\citet{2000ApJ...532.1150Z}:% Zhang and Harding
% $L_\gamma\propto B/P^2$  ( = (spin-down power)$^{1\over2}$), 
%\citet{2004IAUS..218..105L}: %Lorimer
%$B^2\propto(P Pdot)$,$L_{radio}\propto B^2$, $B\propto e^{-t/\tau}$.
% \citet{2002ApJ...565..482G}:%Gonthier 2002
% assume B=const.
% Radio $N(L)=L^{-1}$: can relate directly to $L_\gamma$ using these relations?.
According to the polar-cap model of
 \citet{2002ApJ...565..482G} %Gonthier 2002:
%$P=(A+Bt)^{1\over2}$, 
, $L_\gamma\propto P^{-x} $, x= 27/14 or 9/4 depending on spin-down luminosity.
where the proportionality constant depends roughly linearly on the  magnetic field  and the cosine of the  angle between rotation and dipole axes.

% $dN/dL=dN/dt\ dt/dL\propto dt/dL$, $dL/dt\propto (A+Bt)^{-{x\over2}-1}  \propto L_\gamma^{{-{x\over2}-1}/-{x\over2}}  \propto L_\gamma^{-{2+ x\over x}} $

The luminosity function can be then estimated as follows:
$dN/dL=dN/dP\ dP/dL = dN/dt\ dt/dP\ dp/dL$
$ = P\ dP/dL= P\ P^{1+x} = P^{2+x} = L^{-{2+ x\over x}} $
since $dP/dt=P^{-1}$.
Hence  $N(L)\propto L^{-2.03,-1.89}$, for x=27/14, 9/4.
so a luminosity function index -2 is sufficient to approximate this model.
Obviously distributions of $P_o$, $B_{12}$ and $\alpha$ will modify this simple law,
but we anyway allow for a much wider range of indices in our synthesis since we want
 to avoid dependence on a particular source model.
The outer gap model predicts  $L_\gamma\propto P^{-2/7} $
 (Zhang,Zhang and Cheng 2000) so   $N(L)\propto L^{-8}$
which is not bracketed by our range of models.
But in new outer gap model Zhang et al 2004 604,317 population synthesis gives
 $L_\gamma\propto P^{-1.5} to  P^{-2} $ (i.e. $Edotsd^{.38, .46,1}$ depending on the sample.
This is again covered by our grid of models.
}

\subsection{Known populations}
%%%%%%%%%%%%%%%%%%%%%%%%%%%%%%%%%%%%%%%%%%%%%%%%%%%%%%%%%%%%%%%%%

We start with the EGRET catalogue sources (excluding AGN identifications),
 attempt to reproduce their source counts by population synthesis
and hence estimate the contribution of unseen members of this population to the diffuse emission.
The luminosity function is assumed to be a power law, the index and limits being  free parameters.
In our reference model the range of luminosities considered is  $\Lg$($>$100 MeV) = $10^{36} - 10^{39}$ s$^{-1}$,
covering the range of detected pulsars as discussed in the Introduction.
%We also consider minimum luminosities   L($>$100 MeV) = $10^{35} , 10^{37}$ s$^{-1}$ since this is a major uncertainty.
The local density $\rho$ is  fixed by the requirement that  the low-latitude source counts are reproduced.
%The synthesized models illustrated particular choice of parameters, but can be scaled to
%other luminosities and densities as explained in Section 2.
%(NB Isotropic luminosity since in the beam, and density of pulsars pointing to us).
% The density is constrained from both low-latitude and high-latitude source counts.

For EGRET we use a limiting flux ($>$100 MeV) of  $ 10^{-7}$ cm$^{-2}$ s$^{-1}$; fainter sources are detected by EGRET at high latitudes
but in the plane the limit is higher: 
%(Gonthier 2002  table 5:
  $1.6\ 10^{-7}$ $|b|<10^o$, $0.7\ 10^{-7}$  $|b|>10^o$).
For  a luminosity function index -1.5, 
%(as expected for pulsars e.g. in the outer gap  model),
$\rho=37$ kpc$^{-3}$ (model 1b), the low-latitude source counts in \regionH (hereafter region H) are reproduced (37 above threshold), the
high-latitude sources are very few (so they must be extragalactic in this case),
and about  6\% of the Galactic emission 
( $2\ 10^{-4}$ \fluxunits ) in this region
comes from the 4000 sources below the threshold.
The fluxes above and below the threshold are about equal, $1.5,1.2\ 10^{-5}$ \fluxunits $>$100 MeV in the simulation,
 compared to   $1.4\ 10^{-5}$ \fluxunits for the 37 EGRET sources in region H.
%The longitude profile (Fig xx) shows that the sources below the threshold cause significant fluctuations
%which would warrant a more in-depth analysis of the EGRET data to test this model.
%This is an example of a plausible scenario, fully consistent with the observations but by no means unique.

For a luminosity function index -2.0,
% (as expected for pulsars in the polar cap model),
model 1c,
 again  choosing   $\rho$ so that the low-latitude source counts in region H are reproduced,
13\% of the Galactic emission comes from the 26000 sources below the threshold.

How critical is the luminosity function shape ?
Flattening the luminosity function index from -1.5 to -1.0, 
 about  4\% of the Galactic emission comes from the 728 sources below the threshold.
Steepening the luminosity function index to -2.5, 
 about  28\% of the Galactic emission comes from the   $10^5$    sources below the threshold.
So even a major difference in assumed luminosity function shape does not change the conclusion that a  significant
contribution to the diffuse emission must come from sources physically like those in the EGRET catalogue but below the detection threshold.
A steeper luminosity function or a lower minimum luminosity leads to a larger  contribution from unresolved sources.

The high-latitude source counts are not very constraining, due to the low space density for these high luminosity sources,
but the predicted counts are consistent with (i.e. do not exceed)   the observed counts.
For a steep luminosity function (index -2.0) or a low minimum luminosity ($10^{35}$~\Lgunits~), a significant fraction of the high-latitude sources can be Galactic.
With the small statistics a quantitative comparison of observed and predicted shape of N(S) is difficult,
any index considered here reproduces the observed counts reasonably,
and no distinction between the various indices can be made.

%NB All-sky N(S) could be useful here, seems to give good agreement.

%Other potentially viable cases can be excluded:
%for luminosity index -1.5, increasing to  $\rho=370$ the entire high-latitude (non-AGN-identified)
% source counts are reproduced as Galactic sources,
%but the intensity from sources in the plane is then close to the total observed diffuse emission and there are a factor 10 too many
%low-latitude sources predicted;
%so  for this luminosity index we cannot  have Galactic sources as the high-latitude unidentified sources.
%For  luminosity functions steeper than -2 on the other hand,  this situation {\it is} possible (case 1d)
%without violating the observed diffuse emission or the low-latitude source counts.

%%%%%%%%%%%%%%%%%%%%%%%%%%%%%%%%%%%%%%%%%%%%%%%%%%%%%%%%%%%%%%%%%%%%%%%%%%%%
  \begin{figure*}        % *=2 columns
  \centering    
 \includegraphics[width= 6cm]{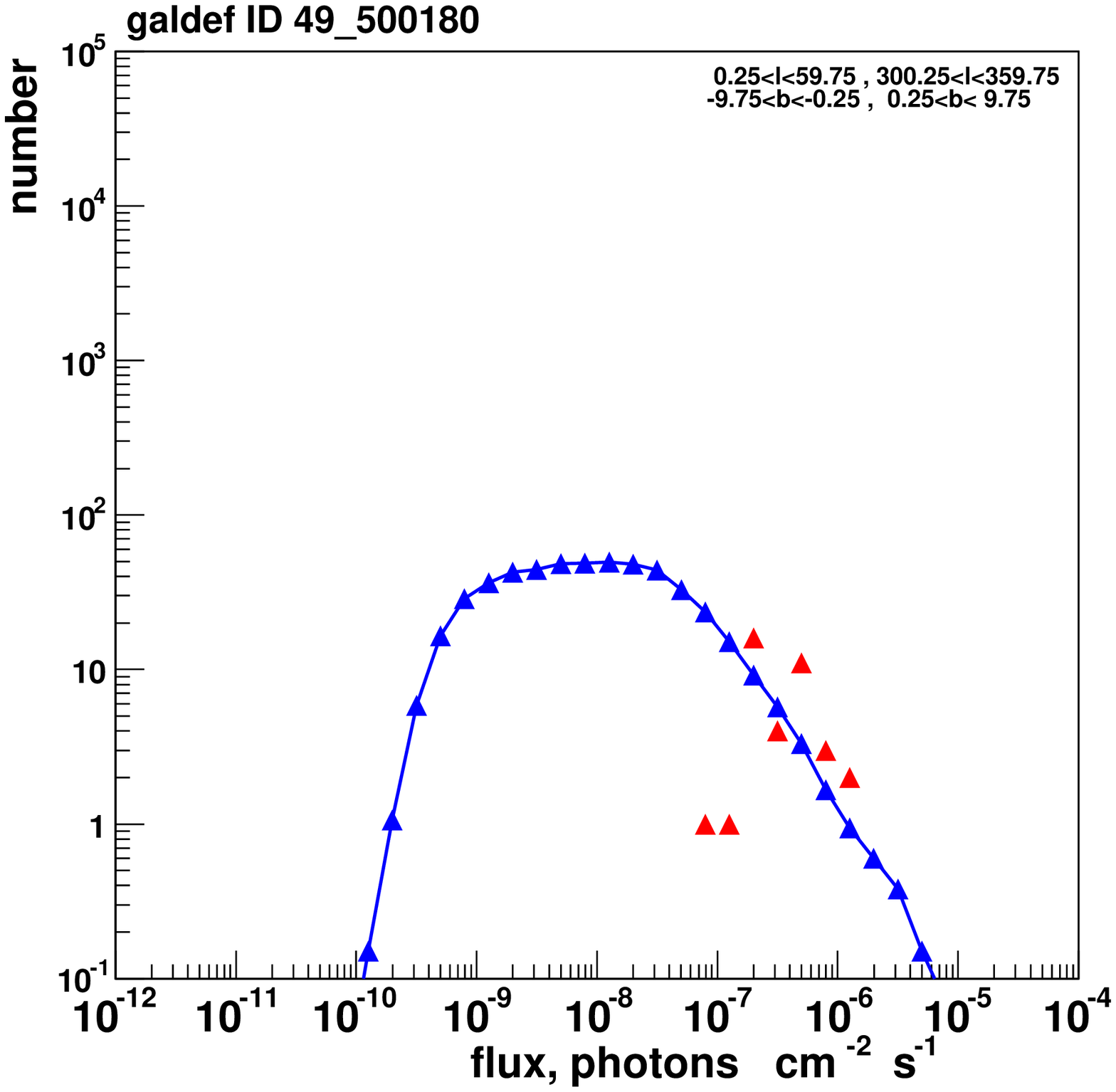}
 \includegraphics[width= 6cm]{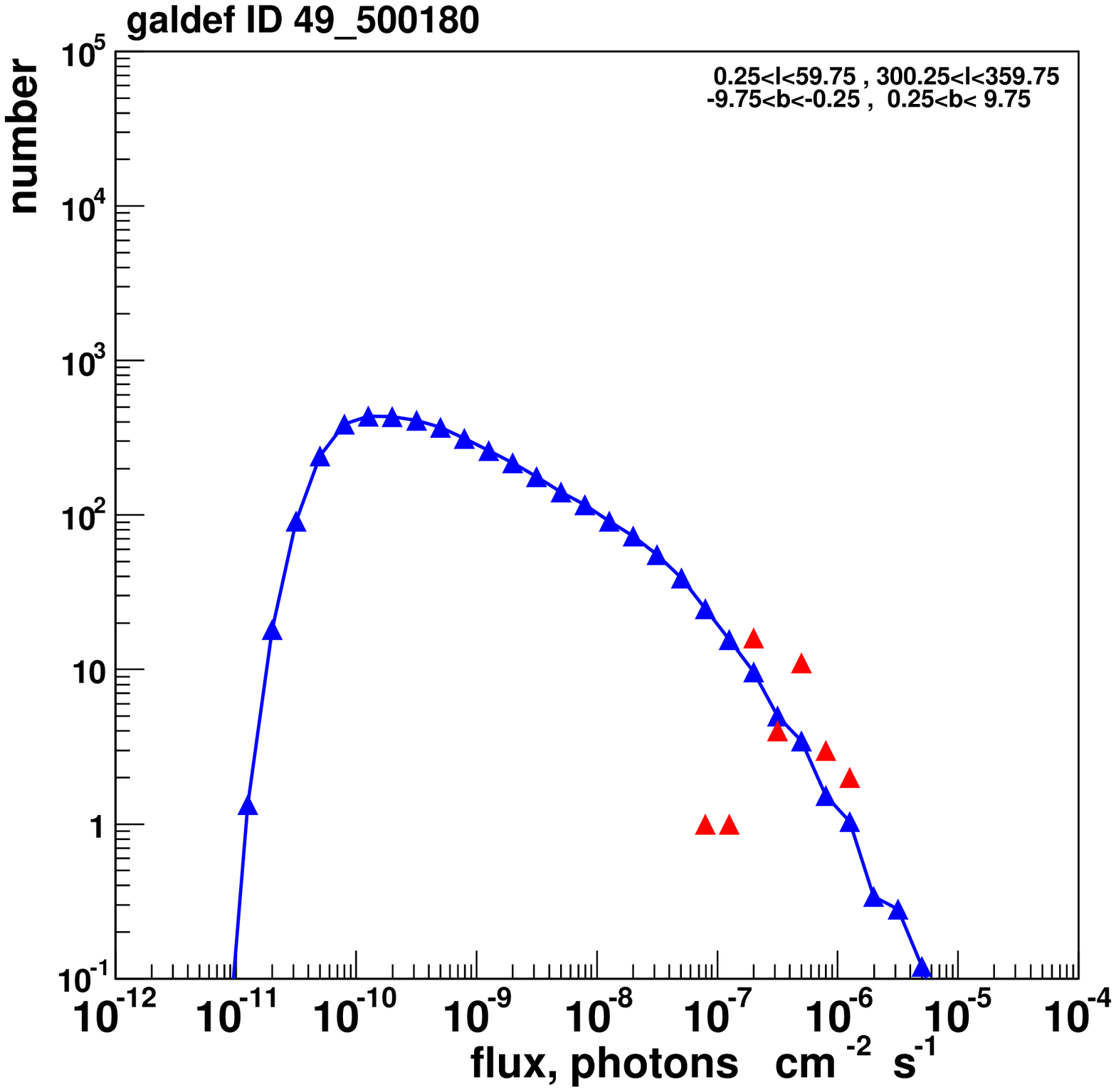}
 \includegraphics[width= 6cm]{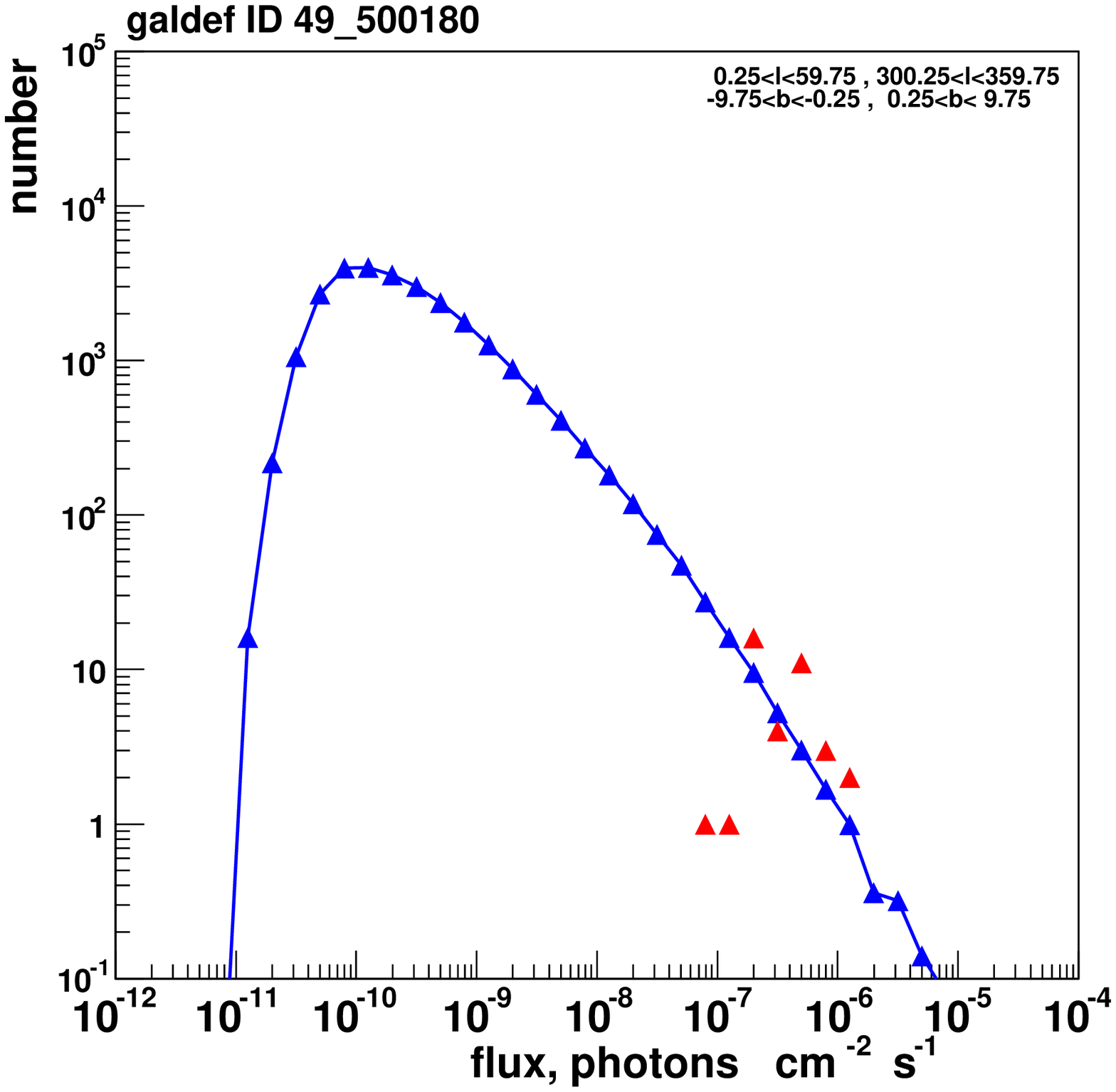}
 \includegraphics[width= 6cm]{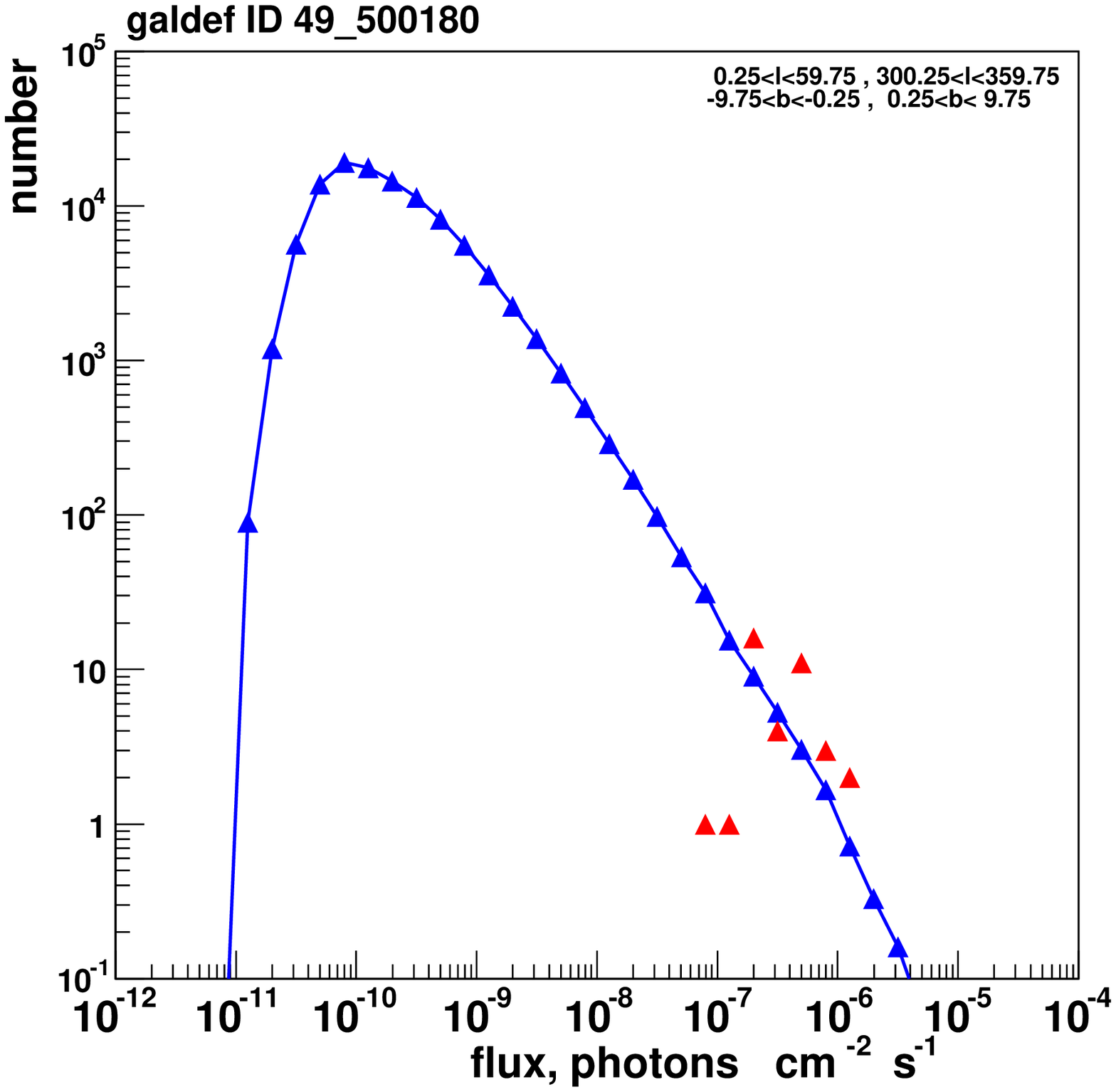}
  \caption{  Differential source counts in \regionH (region H). Blue, connected points:  models 1a-d,           
             luminosity index -1.0,-1.5,-2,-2.5 (left to right, top to bottom).
    Red unconnected points: 3rd EGRET Catalogue, excluding AGN identifications.   
   }
  \label{NS_1a}
  \end{figure*}
%%%%%%%%%%%%%%%%%%%%%%%%%%%%%%%%%%%%%%%%%%%%%%%%%%%%%%%%%%%%%%%%%%%%%%%%%%%%
%%%%%%%%%%%%%%%%%%%%%%%%%%%%%%%%%%%%%%%%%%%%%%%%%%%%%%%%%%%%%%%%%%%%%%%%%%%%
%  \begin{figure*}
%  \centering    
%  \includegraphics[width= 8cm]{NS_1ahilat.eps}
%  \includegraphics[width= 8cm]{NS_1bhilat.eps}
%  \includegraphics[width= 8cm]{NS_1chilat.eps}
%  \includegraphics[width= 8cm]{NS_1dhilat.eps}
%  \caption{Source counts at high latitudes for models 1a-d (left to right, top to bottom).}
%  \label{NS_hilat}
%  \end{figure*}
%%%%%%%%%%%%%%%%%%%%%%%%%%%%%%%%%%%%%%%%%%%%%%%%%%%%%%%%%%%%%%%%%%%%%%%%%%%%

%%%%%%%%%%%%%%%%%%%%%%%%%%%%%%%%%%%%%%%%%%%%%%%%%%%%%%%%%%%%%%%%%%%%%%%%%%%%
  \begin{figure*}
  \centering    
 \includegraphics[width= 6cm]{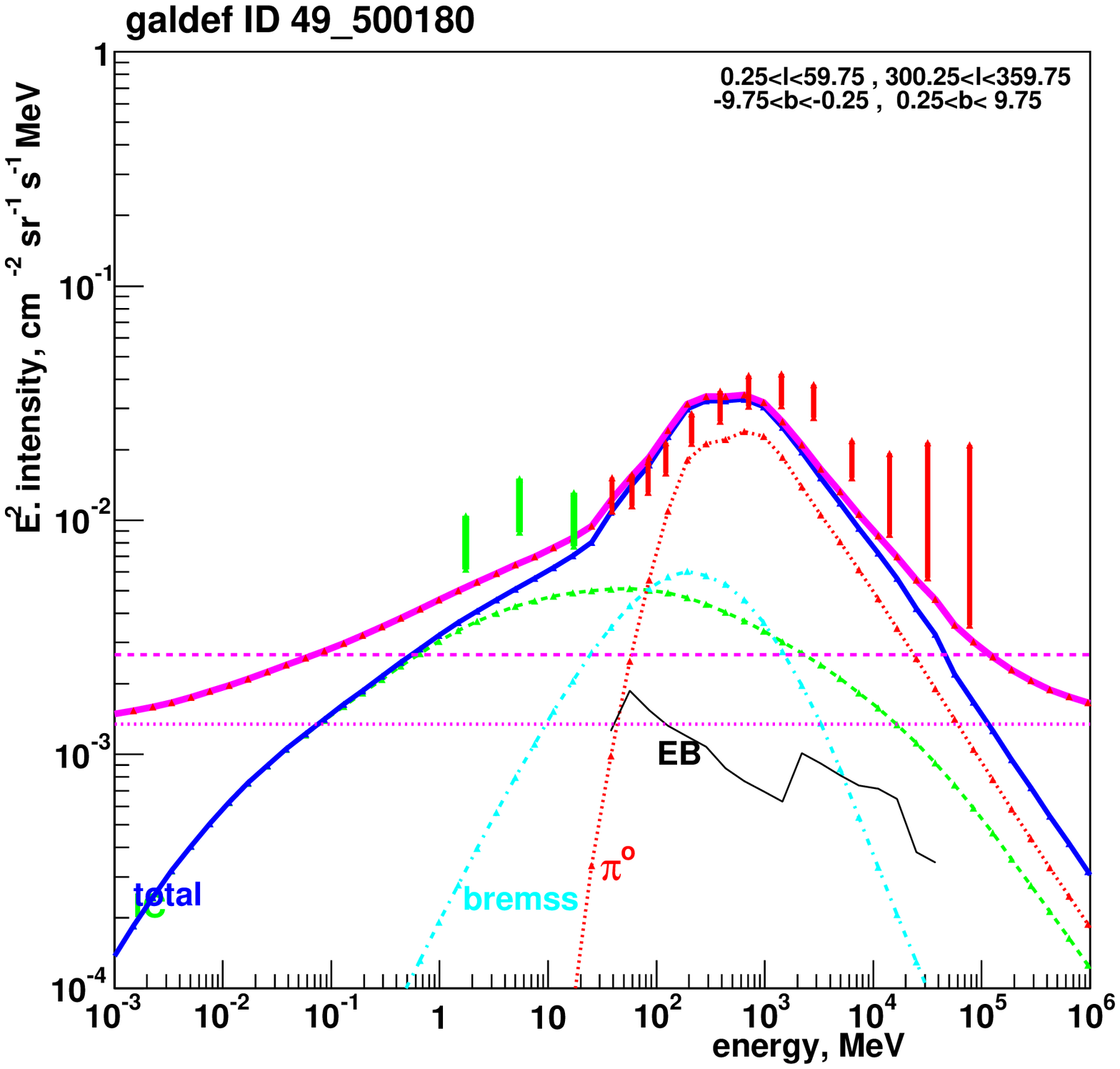}
 \includegraphics[width= 6cm]{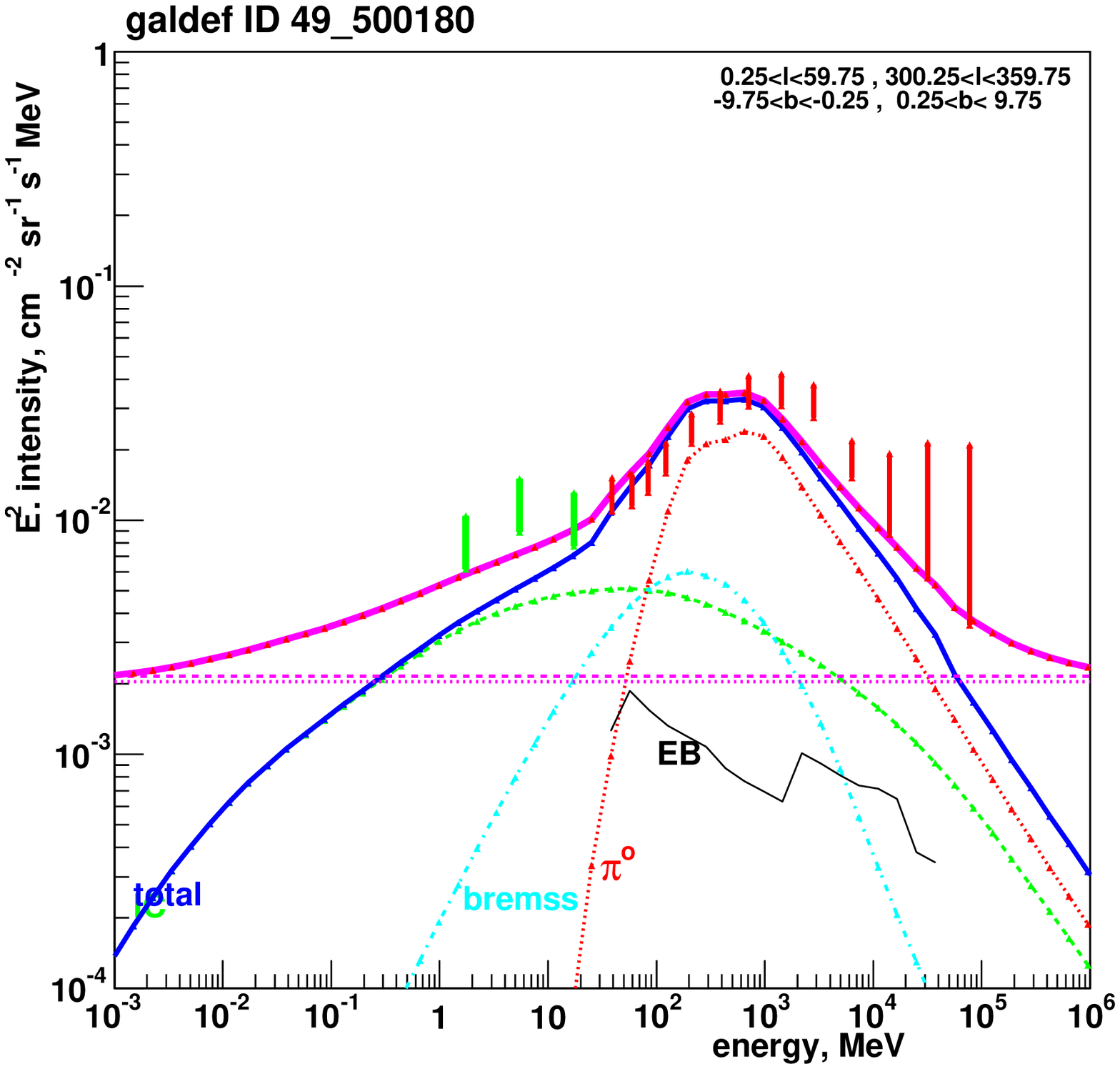}
 \includegraphics[width= 6cm]{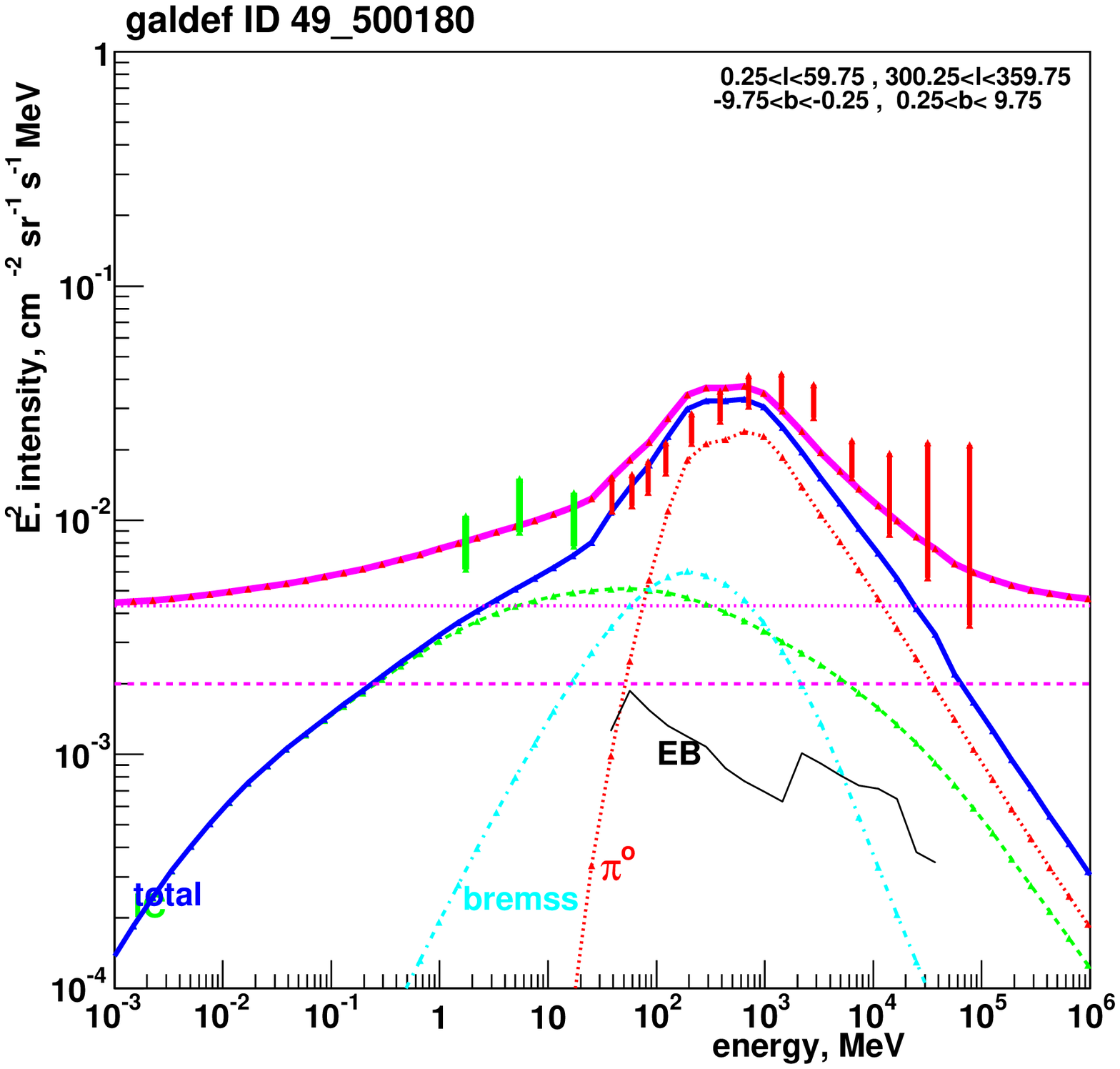}
 \includegraphics[width= 6cm]{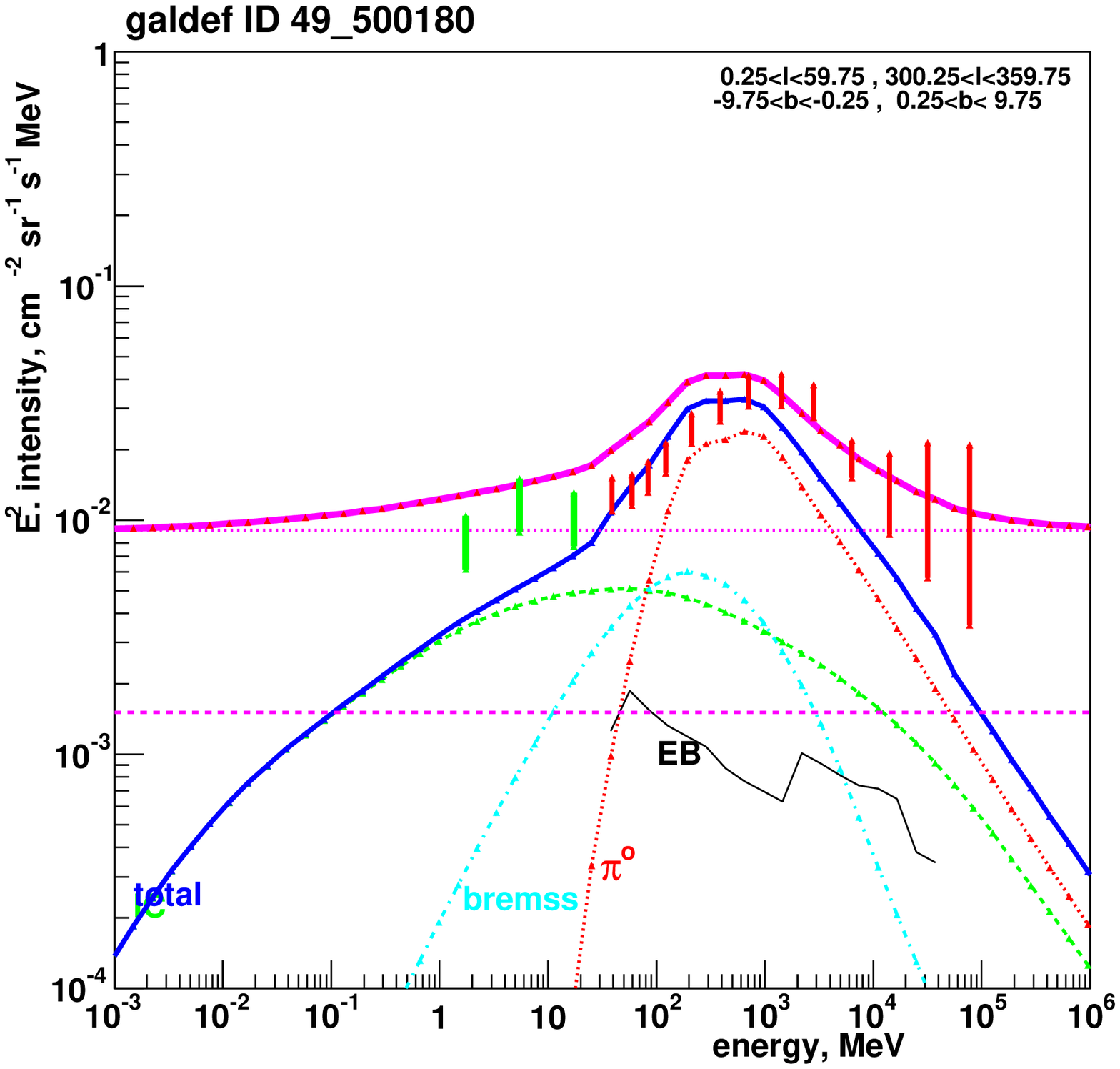}
  \caption{Spectra in region H (\regionH)  for models 1a-d.             
            luminosity index -1.0,-1.5,-2,-2.5 (left to right, top to bottom).
   Sources below (dotted, cyan) and above (dashed, cyan) the EGRET detection limit are also shown
 together with sources below the limit  added to the conventional interstellar model from
   \citet{2004ApJ...613..962S}  %SMR2004
 (continuous, cyan).
   Data: EGRET, COMPTEL.
           }
  \label{spectra}
  \end{figure*}
%%%%%%%%%%%%%%%%%%%%%%%%%%%%%%%%%%%%%%%%%%%%%%%%%%%%%%%%%%%%%%%%%%%%%%%%%%%%

%%%%%%%%%%%%%%%%%%%%%%%%%%%%%%%%%%%%%%%%%%%%%%%%%%%%%%%%%%%%%%%%%%%%%%%%%%%%
%  \begin{figure*}
%  \centering    
%  \includegraphics[width= 6cm]{NS_2H.eps}
%  \includegraphics[width= 6cm]{NS_2hilat.eps}
%  \includegraphics[width= 6cm]{spectrum_2.eps}
%  \caption{Source counts in region H (left) and high latitudes (right ), and spectra,  for model 2 }
%  \label{NS_1b}
%  \end{figure*}
%%%%%%%%%%%%%%%%%%%%%%%%%%%%%%%%%%%%%%%%%%%%%%%%%%%%%%%%%%%%%%%%%%%%%%%%%%%%

%%%%%%%%%%%%%%%%%%%%%%%%%%%%%%%%%%%%%%%%%%%%%%%%%%%%%%%%%%%%%%%%%%%%%%%%%%%%
%  \begin{figure*}
%  \centering    
%  \includegraphics[width= 8cm]{NS_2H.eps}
%  \includegraphics[width= 8cm]{NS_2hilat.eps}
%  \includegraphics[width= 8cm]{spectrum_3.eps}
%  \caption{Source counts in region H (left) and high latitudes (right ), and spectra,  for model 3 }
%  \label{NS_1b}
%  \end{figure*}
%%%%%%%%%%%%%%%%%%%%%%%%%%%%%%%%%%%%%%%%%%%%%%%%%%%%%%%%%%%%%%%%%%%%%%%%%%%%

%%%%%%%%%%%%%%%%%%%%%%%%%%%%%%%%%%%%%%%%%%%%%%%%%%%%%%%%%%%%%%%%%%%%%%%%%%%%
%  \begin{figure*}
%  \centering    
%  \includegraphics[width= 8cm]{NS_4H.eps}
%  \includegraphics[width= 8cm]{NS_4hilat.eps}
%  \includegraphics[width= 8cm]{spectrum_4.eps}
%  \caption{Source counts in region H (left) and high latitudes (right ), and spectra,  for model 4 }
%  \label{NS_1b}
%  \end{figure*}
%%%%%%%%%%%%%%%%%%%%%%%%%%%%%%%%%%%%%%%%%%%%%%%%%%%%%%%%%%%%%%%%%%%%%%%%%%%%

%%%%%%%%%%%%%%%%%%%%%%%%%%%%%%%%%%%%%%%%%%%%%%%%%%%%%%%%%%%%%%%%%%%%%%%%%%%
\subsection{Unseen / dim populations}
%%%%%%%%%%%%%%%%%%%%%%%%%%%%%%%%%%%%%%%%%%%%%%%%%%%%%%%%%%%%%%%%%%%%%%%%%%%
We turn now to source populations mainly below the detection threshold,
with the aim of placing constraints on their properties. The relevant models are
2, 3 and 4 in Table 1.

%We start with a  luminosity function is assumed to be a power law with index -1.5 from L($>$100 MeV) = $10^{35} - 10^{37}$ s$^{-1}$.    RUN THIS CASE
%For $\rho=1000$,  30\% of the diffuse emission in region H is from the 10$^5$ sources below the threshold. Only 2 are above threshold.
%At high latitudes there are 10 sources above the limit, compatible with being a subset of the 98 EGRET unidentified high-latitude sources.

The  luminosity function is assumed to be a power law with index -1.5 and various luminosity ranges.
The normalization is chosen to give about a tenth of the EGRET diffuse emission, as an illustration of a plausible level.
Consider first  $\Lg$($>$100 MeV) = $10^{36} - 10^{37}$ \Lgunits, model 2.
For $\rho= 500$,   15\% of the diffuse emission in region H is from the $5\ 10^4$ sources below the threshold and 44 of the  EGRET unidentified high-latitude sources
 are Galactic
and 17 (cf 3rd EGRET Catalogue: 37) sources are above the limit in region H. 
Longitude profiles (not shown here) indicate that the sources below the threshold cause fluctuations in the emission
 which however are not distinguishable from the interstellar emission.
For  $\Lg$($>$100 MeV) = $10^{35} - 10^{36}$ \Lgunits,   $\rho=2\ 10^3$, (model 3),  7\% of the diffuse emission in region H is from the $2\ 10^5$ sources below the threshold,
  2 sources are above threshold in region H, and 19 at high latitudes.
For  $\Lg$($>$100 MeV) = $10^{34} - 10^{35}$ \Lgunits, $\rho=2\ 10^4$, (model 4),  7\% of the diffuse emission in region H is from the $2\ 10^6$ sources below the threshold,
 no sources are above the threshold in region H, and  only 5 at high latitudes.
It follows that for  $\Lg$($>$100 MeV) $<10^{35}$ \Lgunits all  of the `diffuse' emission {\it could} come
 from sources without violating source counts anywhere on the sky (e.g. by scaling up $\rho$ in model 4).
 This is of course highly unlikely given our knowledge of interstellar emission processes but cannot be excluded from \gray data alone.
%For higher luminosities this situation {\it may} be excluded by high-latitude source counts as shown below.

%For the high-latitude counts we can consider the extreme cases that those without  firm AGN identifications are all either (1)  extragalactic or (2) Galactic.

If we consider that all the high-latitude unidentified sources are Galactic then a dim dense population like in models 2,3,4 is 
{\it necessary} in addition to the bright population. Strict limits are then set by the requirement of not violating
the observed  diffuse emission considering plausible interstellar emission.
If instead they are extragalactic, the dim Galactic population is not required.

\begin{table*}    % *=2 columns
\label{models}      % is used to refer this table in the text
\caption{Summary of population synthesis models. Using thresholds $S_{EGRET}=1\ 10^{-7}$ cm$^{-2}$ s$^{-1}$, $S_{GLAST}=4\ 10^{-9}$ cm$^{-2}$ s$^{-1}$.
For comparison, the EGRET measured diffuse emission in region H (\regionH) is $2\ 10^{-4}$ \fluxunits}
\centering                          % used for centering table
\begin{tabular}{l l l l l l l l l l l}        % left and centered columns (11 columns)
\hline\hline                 
Model&        & $L_{min}$ &   $L_{max}$      & $\alpha$ &$\rho(R_\odot)$ &$N(>S_{EGRET})$ /                 &$N(>S_{EGRET})$/ &$F(>S_{EGRET})$ /&$N(>S_{GLAST})$ / &$F(>S_{GLAST})$ \\   
     &        &           &                  &          &            &        $N(S<S_{EGRET})$   &$N(<S_{EGRET})$   &$F(S<_{EGRET})$            &$N(S<S_{GLAST})$  &$F(<S_{GLAST})$ \\   
     &      &   $>$100    & ph s$^{-1}$ &          &kpc$^{-3}$     & Region H               & high lat.     &region H                     &region H     \\       
     &      &      MeV    &             &          &               &                        &               &          $10^{-4}$ cm$^{-2}$ s$^{-1}$     \\
\hline     
\hline              
  EGRET 3EG   &  &  -    &         -          &    -    &   -  &  37/1           & 47/70  &     0.14 / .01       \\
\hline
\hline
% 1.1a &         & $10^{37}$&  $10^{39}$         & -1.0    &  4.8 &37.2/471         &   /    &     0.18 /0.07 & 332/176  &0.25/0        \\
% 1.1b &         & $10^{37}$&  $10^{39}$         & -1.5    &  10  &  34/1015        &1.3/0   &     0.09 /0.09 & 465/584  &0.22/0.01     \\
% 1.1c &         & $10^{37}$&  $10^{39}$         & -2.0    &  30  &  34/3090        &  8/0   &     0.07 /0.17 &945/2247&0.28/0.03 \\
% 1.1d &         & $10^{37}$&  $10^{39}$         & -2.5    &  40  &  30/4134        &  4/0   &     0.07 /0.16 &915/4253 &0.22/0.04\\
%\hline
%\hline
 1a &         & $10^{36}$&  $10^{39}$         & -1.0    &  7.3 &  37.5/728       &1.4/0.5 &     0.16 /0.08 & 339/426  &0.23/0     \\ 
 1b &         & $10^{36}$&  $10^{39}$         & -1.5    &  37  &  37/3899        &4.2/4.0 &     0.15 /0.12 & 578/3357 &0.24/0.01  \\  
 1c &         & $10^{36}$&  $10^{39}$         & -2.0    & 250  &  37/26550       & 18/39  &     0.14 /0.26 &1167/25420  &0.29/0.10  \\
 1d &         & $10^{36}$&  $10^{39}$         & -2.5    &1000  &  36/106370      & 59/177 &     0.11 /0.56 &2003/104400  &0.34/0.33  \\
\hline
%\hline
% 1.3a &psr      & $10^{35}$&  $10^{39}$         & -1.0    & 10   &  37/1024        &1.2/1.1 &     0.16 /0.08 & 346/715  &0.24/0.00  \\  
% 1.3b &psr      & $10^{35}$&  $10^{39}$         & -1.5    & 120  &  37/12718       &5.1/23  &     0.15 /0.12 & 584/12171&0.25/0.03  \\  
% 1.3c &psr      & $10^{35}$&  $10^{39}$         & -2.0    &2350  &  37/$2.5\ 10^5$    & 31/520 &     0.12 /0.37 &1264/$2.5\ 10^5$   &0.29/0.20  \\
%\hline
%\hline         
  2 &          & $10^{36}$&  $10^{37}$         & -1.5    &$500 $&  17/$0.5\ 10^5$ & 44/74  &     0.06 /0.32 &1138/$5\ 10^4$&0.16/0.20 \\   
%\hline 
  3 &          & $10^{35}$&  $10^{36}$         & -1.5    &$2\ 10^3$&2/$2  \ 10^5$ & 19/490 &     0.001/0.14 &224/$2\ 10^5$&0.02/0.12\\ 
%\hline
  4 &          & $10^{34}$&  $10^{35}$         & -1.5 &$2\ 10^4$&   0/$2  \ 10^6$ &  5/4728 &    0.0  /0.14 &94/$2\ 10^6$ &0.007/.14    \\  
\hline
\hline
%  5 &  stars   & $10^{31}$&  $10^{32}$         & -1.5    &$10^8$&  0 /$10^{10}$   &  0/$2\ 10^8$&   0.0/0.73      \\ % scaling from 4
%\hline
\end{tabular}
\end{table*}
\section{Comparison with physical pulsar population synthesis}
%%%%%%%%%%%%%%%%%%%%%%%%%%%%%%%%%%%%%%%%%%%%%%%%%%%%%%%%%%%%%%%%%%%%%%%%%%%%%%%%%%%%%%

It is interesting to see how our simple generic approach matches detailed specific models.
In their  pulsar population synthesis based on their polar cap model,
\citet{2004ApJ...604..775G} % Gonthier et al.
find 26 pulsars detectable by EGRET, which presumably would mean a substantial fraction of the unidentified sources are pulsars.
They  predict that 600 pulsars will be detectable by GLAST for a threshold   $2-5\ 10^{-9}$ cm$^{-2}$ s$^{-1}$.
%This matches best our model 1b,c or d, but not 1a   (predicts less).
This matches best our  model 1b (index -1.5)

% with  578 sources above  $4\ 10^{-9}$ cm$^{-2}$ s$^{-1}$ in region H.
%Steeper luminosity functions do not increase the source count above this threshold.
%As with our  model, theirs is adjusted to give approximate agreement with the EGRET detected pulsars.
%Their model is therefore equivalent to a luminosity function index equal to, or steeper, than -1.5.
%NB can get values from N($>S$) table, maybe add to table 2.

The outer gap model of
\citet{2000A&A...357..957Z} % Zhang et al 2000 AA357,957, 
 predicts 32  pulsars detectable by EGRET, which again would  mean a substantial fraction of the unidentified sources are pulsars.
%(and they suggest this explains a statistical association of the sources with SNR and OB associations,  having a common origin with the pulsars).
This model   predicts 1180 GLAST pulsars,
for assumed threshold of  $4\ 10^{-9}$ cm$^{-2}$ s$^{-1}$
  matching best our model 1c  with luminosity function index -2.0.

%(see also Table 1 of Harding AGILE Review).

%%%%%%%%%%%%%%%%%%%%%%%%%%%%%%%%%%%%%%%%%%%%%%%%%%%%%%%%%%%%%%%%%%%%%%%%%%%%%%%%%%%%%%%%%%%%%%%
\section{Sources can produce the MeV, GeV excesses}
%%%%%%%%%%%%%%%%%%%%%%%%%%%%%%%%%%%%%%%%%%%%%%%%%%%%%%%%%%%%%%%%%%%%%%%%%%%%%%%%%%%%%%%%%%%%%%%
The unresolved source fraction is energy-dependent, so that one can ask whether it can
produce the well-known GeV excess over the standard cosmic-ray interaction models.
Consider first  model 1c (Fig 3c) with  spectral index -2.0 and  luminosity function index -2.0,
combined with the `conventional' interstellar emission model from
\citet{2004ApJ...613..962S},  %SMR2004
but with the cosmic-ray source distribution from
\citet{2004A&A...422L..47S}. %Strong 
The GeV excess is {\it not} reproduced. 
However for this source spectrum  the source contribution to the COMPTEL (1--30 MeV) and INTEGRAL (.02 -- 1 MeV) ranges
 might  provide an explanation of the excess above the interstellar emission at those energies.
Sources with a  Crab-pulsar-like index of -2.1 with a break  above 4 GeV 
 \citep{1998ApJ...494..734F,2001A&A...378..918K} %Fierro, Kuiper
 would be too steep to reproduce the GeV excess, but the COMPTEL diffuse emission  {\it could} be  fitted.% (Fig \ref{spectra_Crab_like}).
% The SPI diffuse ridge measurements
% \citep{2005A&A...444..495S} %Strong
% below 100 keV would however be violated since the pulsar power-law spectrum continues with index -2.3 down to 10 keV 
% \citep{2001A&A...378..918K}. %Kuiper

%\footnote
%{ Crab pulsar not same as SNR, but total: -2.12, peak 1 2.07 peak 2 2.16 (Fierro) so very similar.
% Steepens below 30 MeV (Kuiper) to 2.22. Kuiper Fig 9 is best Crab total pulsed spectrum, with sharp break above 2 GeV.}
%%%%%%%%%%%%%%%%%%%%%%%%%%%%%%%%%%%%%%%%%%%%%%%%%%%%%%%%%%%%%%%%%%%%%%%%%%%%
%  \begin{figure*}
%  \centering    
%  \includegraphics[width= 6cm]{spectrum_g2.1_br4e3.eps}
%  \caption{Sources like the Crab pulsar  {\it cannot}  produce the GeV excess: 
%   spectra for  luminosity index 2.0, spectral index -2.1, break at 4 GeV,
%   to match  Crab pulsar.  $\rho=30$.
%   Sources below detection limit are also shown added to the conventional interstellar model.
%   Data: EGRET, COMPTEL, INTEGRAL/SPI, RXTE.
%   Region A for compatibility with \citet{2005A&A...444..495S}.%Strong etal
%{\it this figure is not latest model: rerun}
%  }
%  \label{spectra_Crab_like}
%  \end{figure*}
%%%%%%%%%%%%%%%%%%%%%%%%%%%%%%%%%%%%%%%%%%%%%%%%%%%%%%%%%%%%%%%%%%%%%%%%%%%%
%%%%%%%%%%%%%%%%%%%%%%%%%%%%%%%%%%%%%%%%%%%%%%%%%%%%%%%%%%%%%%%%%%%%%%%%%%%%
  \begin{figure*}
  \centering    
  \includegraphics[width= 6cm]{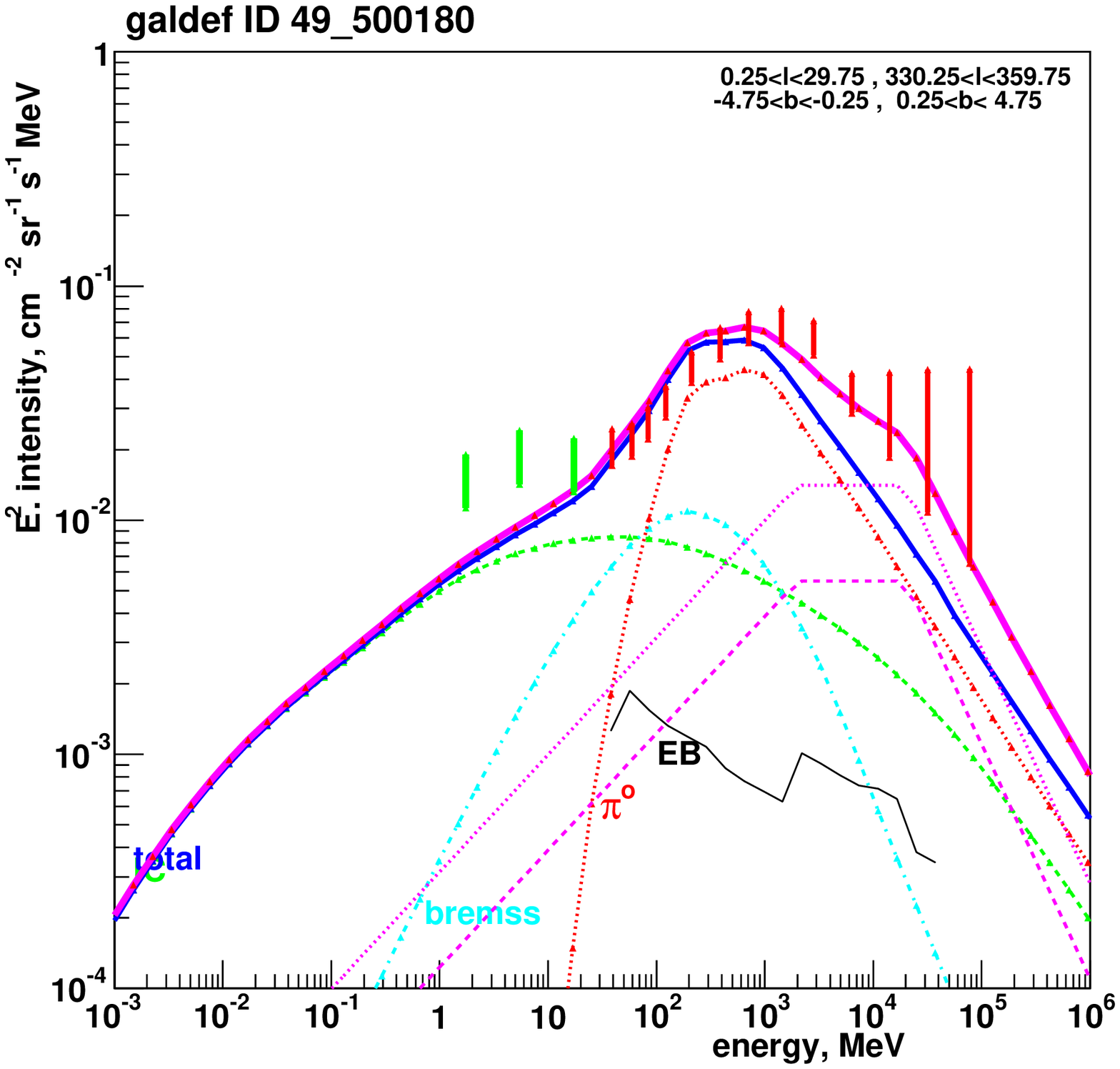}
  \caption{Sources like Geminga, Vela pulsars {\it can} produce the GeV excess: 
   spectra  in \regionA (region A of 
  \citet{2004ApJ...613..962S}  %SMR2004
  ) for  L($>$100 MeV) = $10^{36} - 10^{39}$ s$^{-1}$. luminosity index 2.0, spectral index -1.5, break at 2 GeV to -2.0
   to match  Geminga, Vela. $\rho=150$.
    Sources below (dotted, cyan) and above (dashed, cyan) the EGRET detection limit are also shown
 together with sources below the limit  added to the conventional interstellar model from
   \citet{2004ApJ...613..962S}  %SMR2004
 (continuous, cyan).
   Data: EGRET, COMPTEL.
      }
  \label{spectra_Geminga_like}
  \end{figure*}
%%%%%%%%%%%%%%%%%%%%%%%%%%%%%%%%%%%%%%%%%%%%%%%%%%%%%%%%%%%%%%%%%%%%%%%%%%%%

Consider now sources with   hard spectra like the Geminga (index -1.42, 30-2000 MeV) and Vela (index -1.62, 30-1000 GeV)
 pulsars  \citep{1998ApJ...494..734F} %Fierro 
(also B1706, B1055  have hard spectra  with a GeV break);
we adopt  a spectral index -1.5, with a break at 2 GeV to -2.0,   luminosity index 2.0, $\rho=30$ (cf.  model 1c);
 the GeV excess is  easily produced (Fig \ref{spectra_Geminga_like}).
Indeed these pulsars show  maxima in $E^2\ F(E)$ around 2 GeV very reminiscent of the Galactic GeV excess. % (Fierro Figs 8,10). 
This explanation of the GeV excess  was also proposed by
\citet{1998MNRAS.301..841Z} % Zhang and Cheng
on the basis of their outer-gap pulsar model,
but our result is not dependent on a particular physical model.

However the diffuse emission  below 30 MeV is not explained by such sources.
It is then tempting to propose a mixture of Crab-like and Geminga/Vela-like pulsars to produce both the MeV and GeV excesses
and reproduce the entire ridge spectrum.

The spectra of the unidentified low-latitude sources scatters broadly around -2 (-1.7 to -2.7) 
\citep{2000A&A...357..957Z} %Zhang Zhang and Cheng 2000
but the details of the spectra are not sufficient to determine whether this supports our hypothesis about the GeV excess. GLAST will contribute
significantly on this point.

%%%%%%%%%%%%%%%%%%%%%%%%%%%%%%%%%%%%%%%%%%%%
\section{Will GLAST  resolve the issue ?}
%%%%%%%%%%%%%%%%%%%%%%%%%%%%%%%%%%%%%%%%%%%%
Although GLAST will not detect all the \gray sources in the Galaxy, it will resolve essentially all the source {\it flux}
of the bright populations. Thus for model 1c (luminosity index -2.0), 75\% of the source flux is above a threshold  $4\ 10^{-9}$ cm$^{-2}$ s$^{-1}$,
compared to 35\% for the EGRET threshold. For model 1b (luminosity index -1.5), 96\% of the source flux is above the GLAST threshold,
 compared to 55\% for the EGRET threshold.
For the dimmer populations  ($\Lg$($>$100 MeV) $<10^{37}$ \Lgunits) progressively less of the flux will be resolved.
Even for $\Lg$($>$100 MeV) $=10^{36}-10^{37}$ \Lgunits (model 2)  only 44\% of the flux is above the GLAST threshold,
so the analysis will remain a challenge.

\ifthenelse{0=1}
{
%%%%%%%%%%%%%%%%%%%%%%%%%%%%%
\section{Hard X-ray energies}
%%%%%%%%%%%%%%%%%%%%%%%%%%%%%
Although not the main focus of this work, we can apply the same techniques to the hard X-ray range (20-500 keV)
with particular reference to the INTEGRAL results.
\citet{2005A&A...443..485D}  %Dean et al. 
using the  1st IBIS catalogue find luminosities covering the range  L(20-100 keV)= $10^{36-38}$ \Lgergunits =  $10^{43-45}$ \Lgunits,
in particular
LMXB: L(20-100 keV)= $10^{36}- 5\ 10^{37}$ \Lgergunits, HMXB: $3\ 10^{36}- 2\ 10^{37}$ \Lgergunits
or
BH  : L(20-100 keV)= $3\ 10^{35}- 5\ 10^{37}$ \Lgergunits, NS:   $5\ 10^{35}- 3\ 10^{36}$ \Lgergunits.

As before we attempt to reproduce the source counts by population synthesis and then predict the contribution from unresolved sources of the same type,
and assess the possible contribution from dimmer populations.

We use the analysis results of \citet{2005A&A...444..495S}%Strong etal
which give both source and diffuse emission spectra based on INTEGRAL/SPI
using source positions from the 2nd INTEGRAL/ISGRI Catalogue
\citep{2006ApJ...636..765B}.% Bird et al
Review of luminosity index:
\citet{2002A&A...391..923G}% Grimm et al
using RXTE survey data of the Galaxy found index -1.26 for LMXB, -1.64 for HMXB.
 \citet{2005astro.ph.10049S}%Sazonov et al
obtain the luminosity function over a very wide range using RXTE ........
 \citet{2004MNRAS.349..146G} %Gilafanov
using Chandra observations of 11 nearby galaxies found index -1 for LMXB and -1.6 for HMXB.
\citet{2005A&A...443..485D}  %Dean et al.
using the 1st INTEGRAL/ISGRI catalogue with source identifications and distances, give luminosity functions which
have roughly index -1.6 for both LMXB and HMXB, or -1.3 for BH systems,-2.3 for NS systems.
  (see also Dean ISDC presentation).
Using these results as a  guide to the expected range,
we use  luminosity indices -1.0, -1.5 and $\rho=1-1.5$
 which reproduces the source counts; 
 taking a mean spectral index of -3  produces the summed flux from detected sources (Fig X).
Then the 20-50 keV diffuse flux can be explained as unresolved sources, but above this the
diffuse flux is much too hard for these to make a contribution.

A steeper luminosity function with index -1.5 produces too much  20-50 keV diffuse flux and can be excluded for the steep spectrum sources (Fig Y)
but is allowed for the hard spectrum sources.

To produce the spectrum above 50 keV, we need lower luminosity sources to be below the detection threshold, and a hard spectrum;
we use  $10^{41-43}$ \Lgunits, a spectral index -1.5 and a break at 1 MeV, representing AXPs, pulsars or microquasars (see Bosch-Ramon astroph/0601238)
 is used
and a density $\rho=0.5$ or 1.0 depending on luminosity index.
 This population would be invisible to INTEGRAL except for a few of the brightest objects,
and lies below the observed source counts. At the same time it produces the observed diffuse emission in 50--500 keV.

%%%%%%%%%%%%%%%%%%%%%%%%%%%%%%%%%%%%%%%%%%%%%%%%%%%%%%%%%%%%%%%%%%%%%%%%%%%%
%  \begin{figure*}
%  \centering    
%  \includegraphics[width= 6cm]{NS_5.eps}
%  \includegraphics[width= 6cm]{spectrum_5.eps}
%  \caption{INTEGRAL/SPI 18--28 keV source counts in region H  and spectra,  for model 5.
%  Shows that unresolved sources of the high-luminosity population can produce the 20--50 keV diffuse emission.}
%  \label{NS_5}
%  \end{figure*}
%%%%%%%%%%%%%%%%%%%%%%%%%%%%%%%%%%%%%%%%%%%%%%%%%%%%%%%%%%%%%%%%%%%%%%%%%%%%
%%%%%%%%%%%%%%%%%%%%%%%%%%%%%%%%%%%%%%%%%%%%%%%%%%%%%%%%%%%%%%%%%%%%%%%%%%%%
%  \begin{figure*}
%  \centering    
%  \includegraphics[width= 6cm]{NS_6.eps}
%  \includegraphics[width= 6cm]{spectrum_6.eps}
%  \caption{INTEGRAL/SPI 18--28 keV source counts in region H  and spectra,  for model 6.
%  Shows that unresolved sources of the low-luminosity population can produce the 50--500 keV diffuse emission.
%}
%  \label{NS_6}
%  \end{figure*}
%%%%%%%%%%%%%%%%%%%%%%%%%%%%%%%%%%%%%%%%%%%%%%%%%%%%%%%%%%%%%%%%%%%%%%%%%%%%

%%%%%%%%%%%%%%%%%%%%%%%%%%%%%%%%%%%%%%%%%%%%%%%%%%%%%%%%%%%%%%%%%%%%%%%%%%%%%%%%%%%
% 
\begin{table*}    % *=2 columns
\label{models}      % is used to refer this table in the text
\caption{Summary of population synthesis models for hard X-rays. Using $S_{lim}=2\ 10^{-3}$ cm$^{-2}$ s$^{-1}$.
Plotted models indicated by *.
ADD fraction of diffuse ?}
\centering                          % used for centering table
\begin{tabular}{l l l l l l l l l}        % left and centered columns (9 columns)
\hline\hline                 
Model& spatial& $L_{min}$ &   $L_{max}$      & $\alpha$ &$\rho$ &$N(>S_{lim})$ /                 &$N(>S_{lim})$/ &$F(>S_{lim})$ /  \\   
     & distr. &           &                  &          & (R=8.5 kpc)   &        $N(S<_{lim})$   &$N(<_{lim})$   &$F(S<_{lim})$    \\   
     &      &    18-28 keV& photons s$^{-1}$ &          &kpc$^{-3}$     & Region H               & high latitudes&region H  $10^{-4}$ cm$^{-2}$ s$^{-1}$                           \\    % table heading            
\hline     
\hline              
  SPI         &- &  -    &         -          &    -    &   -  &  45/42          &    -   &      4088/454        \\
  catalogue   &  &       &                    &         &      &                 &        &                      \\
\hline
\hline
  5a &pulsars  & $10^{42}$&  $10^{44}$         &  1.0    &  1.5 &  61/95          &   -    &     5263 /526        \\  
  5b &pulsars  & $10^{42}$&  $10^{44}$         &  1.5    &  2.5 &  45/250         &   -    &     4756 /1049       \\ 
  6a &pulsars  & $10^{41}$&  $10^{43}$         &  1.0    &  0.5 &   2/50          &   -    &       81 /104        \\  
  6b &pulsars  & $10^{41}$&  $10^{43}$         &  1.5    &  1.0 &   1/105         &   -    &       76 /119        \\  
  7  &pulsars  & $10^{40}$&  $10^{43}$         &  1.5    &  3.0 &   1/316         &   -    &       68 /117        \\  
\end{tabular}
\end{table*}
%%%%%%%%%%%%%%%%%%%%%%%%%%%%%%%%%%%%%%%%%%%%%%%%%%%%%%%%%%%%%%%%%%%%%%%%%%%%%%%%%%
}{}
%%%%%%%%%%%%%%%%%%%%%%%%%%%%
\section{Conclusions}
%%%%%%%%%%%%%%%%%%%%%%%%%%%%

1. Modelling the contribution from unresolved sources  is  essential to understanding  the diffuse Galactic emission .

\noindent
2. The contribution from unresolved sources to the EGRET low-latitude emission is at least 5-10\%, and can be 20\% for steep luminosity functions.

\noindent
3. An arbitrarily large fraction of the diffuse emission could come from sources  $\Lg$($>$100 MeV) $<10^{35}$ \Lgunits  
 from sources without violating EGRET source counts anywhere on the sky. 

\noindent
4. The GeV excess can be produced naturally by a sufficient population of  sources like Geminga and Vela,
 but this has to be studied with more detailed models.

\noindent
5. Crab-like sources can produce the COMPTEL excess but not the GeV excess.

\noindent
6. A combination of source populations combined with the conventional model of interstellar emission could explain
the full  COMPTEL/EGRET Galactic ridge spectrum.

\noindent
7. Whether  GLAST can settle these issues depends critically on the source luminosity function.

%%%%%%%%%%%%%%%%%%%%%%%%%%%%%%%%%%%
%\begin{acknowledgements}
%\href{http://dx.doi.org/10.1051/0004-6361:20053702}{reference} % does not work
%\url{reference}
%%\href{file:///afs/ipp/u/aws/actions}{notesurl}

%\url{notes}                               %works on iws5 only

%My home page:
%\url{http://www.mpe.mpg.de/~aws/aws.html} %works on c01 and iws5

%A reference:
%\url{http://dx.doi.org/10.1051/0004-6361:20053702}

%About hypertex:
%\url{http://arxiv.org/hypertex}
%\url{http://www.tug.org/applications/hyperref/ftp/doc/manual.html}

% \end{acknowledgements}
%%%%%%%%%%%%%%%%%%%%%%%%%%%%%%%%%%%%%%%%%%%%%%%%%%%%%%%%%%%%%%%

%%%%%%%%%%%%%%%%%%%%%%%%%%%%%%%%%%%
%\bibliographystyle{spmpsci} % from template.tex
\bibliography{strong,gamma_1990-1999,gamma_2000-2006,gamma_extra,astroph,reimer2001,lorimer2004} % bibtex input
\bibliographystyle{aa}

\end{document}